\documentclass[12pt,aps]{revtex4}
\usepackage{epsfig}
\usepackage{amsmath,amssymb,color}
\usepackage{epstopdf}
\parskip=\medskipamount

\newcommand{\be}{\begin{equation}}
\newcommand{\ee}{\end{equation}}
\newcommand{\eq}[1]{(\ref{#1})}
\newcommand{\fig}[1]{Fig.\ref{#1}}

\newcommand\disp{\displaystyle}

\begin{document}

\title{Finite-size effects in exponential random graphs and cluster
evaporation}

\author{A. Gorsky$^{1,2}$ and O. Valba$^{3,4}$}

\affiliation{
$^1$ Institute for Information Transmission Problems of the Russian Academy of Sciences, Moscow,
Russia, \\
$^2$ Moscow Institute of Physics and Technology, Dolgoprudny 141700, Russia,\\
$^3$ Department of Applied Mathematics, National Research University Higher School of Economics,
101000, Moscow, Russia, \\
$^4$ N.N. Semenov Institute of Chemical Physics of the Russian Academy of Sciences,
119991, Moscow, Russia.
}
\begin{abstract}

In this Letter we find numerically the strong finite-size effects in the critical behavior of Erd\H{o}s-R\'enyi (ER) networks  supplemented with  chemical potentials for some motifs, in particular 2-stars and triangles. For the 2-star model 
above the critical value of the chemical potential a ground state looks as star-like graph with the finite set of hubs at ER parameter $p<0.5$ or as the single cluster at $p>0.5$. 
It is found that there exists the critical value of number of nodes $N^{*}(p)$ when the ground state undergoes clear-cut crossover 
and at $N>N^{*}(p)$ the network flows via a cluster evaporation
to the state involving the small star in the ER environment.  The similar evaporation of the cluster takes place 
at $N>N^{*}(p)$ in the Strauss model. We suggest that the entropic trap mechanism is relevant for 
microscopic mechanism behind  the crossover regime. The possible analogies concerning the strong entropic finite-size effects in the holographic description of matrix black hole (BH) formation and evaporation are mentioned.

\end{abstract}
\maketitle
\section{Introduction} 

Exponential random graphs provide examples of the Ising type systems analyzed via the 
conventional methods of the statistical physics and in  general
one can consider canonical, microcanonical and mixed ensembles. These cases correspond
to introducing  chemical potentials for relevant observables,
fixing the values of these observables or combination of the chemical
potentials for part of the observables and fixing the values of the rest. 
They can exhibit the critical
behavior and have a rich phase structure in the space of some control
parameters.

Let us recall the known examples of a critical behavior in 
exponential random graphs.
The network ensemble supplemented with the chemical potential
for the open triads (2-star model)
has been solved in thermodynamic limit
in mean-field approximation in dense regime \cite{2star} and in  sparse regime \cite{alba}. 
The example of the phase transition in the 
exponential model
in the canonical ensemble with the
chemical potential for triangles has been found in \cite{strauss}. The single complete graph 
above the critical value of positive chemical  potential for triangles is formed,
the phenomenon  has been 
explained analytically in \cite{newman,burda}. At a negative chemical
potential for triangles the network becomes bipartite below some
critical value of chemical potential \cite{old1,old2,+maslov}. The critical 
behavior in  microcanonical ensemble with fixed number of links and 
triangles has been considered in \cite{radin1,radin2,radin3}. The mixed 
ensembles with fixed degrees and chemical potential
for triangles has been investigated in \cite{crit}(see,also \cite{foster}
for the previous study). It was discovered
that above the critical value of a chemical potential  the network gets
defragmentated into the fixed number of weakly connected  clusters.

The real networks are finite  hence the role of the finite-size
effects certainly is important for applications. Several aspects 
of the finite-size effects have been elaborated for the scale-free
network where they induced the cut-off in a
degree distribution \cite{ sf1,sf2,sf3,sf4,sf5,sf6}. The finite
size effects influence the  criticality of the epidemic processes
\cite{epid1,epid2} while crossover 
regime at some network size $N^*$ has been found also  in the synchronization
problem in the scale-free networks \cite{synch}. Very recently 
the example of the strong finite-size effect in the peculiar 
phase transition in the microcanonical ensemble of exponential 
random graphs has been found in \cite{radinnew}.

The interesting analysis of the finite-size effects 
in the Anderson model on the random regular graphs (RRG) 
has been performed in \cite{mirlin1,biroli,mirlin2}.
There was the long-standing controversy concerning the ergodicity
of the modes in the delocalized phase and the very existence of the
delocalized non-ergodic phase of the Anderson model on RRG. The careful 
finite N analysis demonstrated that there 
is no  additional phase transition between the ergodic and
non-ergodic delocalized phases  and instead there is the strong
finite-size effect. At some value of $N$ the system undergoes 
the crossover behavior from the near critical regime to the
ergodic delocalized state. A kind of RG analysis has been 
developed in \cite{mirlin2} which supports this picture.

The primary motivation for this study concerns the question about
the influence of the finite-size effects on 
the critical behavior of the exponential random graphs. It 
was usually assumed \cite{old1,old2} on the basis of mean-field
analysis that the transition to the
thermodynamical limit is smooth. We investigated it numerically and it turned out that the 
exponential random graphs undergoes the crossover behavior at some finite $N^{*}(p)$
above the critical value of the chemical potential.
The ground state of the network above the critical chemical
potential looks differently at two sides of crossover. 
For the 2-star model 
above the critical value of the chemical potential the ground state looks as star-like graph of the finite set of hubs at $p<0.5$ or as the single cluster at $p>0.5$. 
It is found that there exists the critical value of number of nodes $N^{*}$ when the ground state undergoes clear-cut crossover 
and at $N>N^{*}$ it flows to the small hub in the ER environment state via a cluster evaporation.  

There are two ways to get the crossover regime.
Following the first strategy it is useful  to choose  the chemical potential for the corresponding
motives well above the critical value and vary N. There is  crossover regime
at some  $N$  when the  the network ground state gets changed.
Another way to capture the crossover regime involves the $N$ dependence
of the critical values of the chemical potentials.
At $N<N^{*}$ critical chemical potential behaves as $N^{-1}$ 
in agreement with the mean-field solutions to the 2-star and Strauss models
while at $N>N^{*}$ their $N$-dependence changes drastically. Let us emphasize 
that the ground state above $N^{*}$ involves several components with the different
mean degrees and can not be captured via the naive mean-field analysis.

We shall briefly discuss the possible microscopic description of the cluster
evaporation which is responsible for the crossover regime. Due to the 
non-homogeneity of the network there are the entropy  gradients and therefore
the entropic forces. We argue that the entropic trap could be one of the
mechanisms responsible for the cluster evaporation. The emerging clusters
in the matrix models provide the toy model for the black holes formed
from the D0 particles in the holographic framework \cite{hanada,cotler}. Our findings concerning the
role of $\frac{1}{N}$ effects in the derivation of the ground state
could be of some use for the microscopic analysis of the black hole formation
and decay within the matrix model framework.

The paper is organized as follows. In Section 2 we consider the different aspects 
of the crossover behavior at some critical value $N^{*}$. In Section 3 we discuss the possible
microscopic explanation of crossover and comment on the gravity analogies.
Some open question are formulated in Conclusion.

\section{Finite size effects and the critical behavior of 2-stars and Strauss models}

\subsection{The models}

We define an ensemble consisting of the set of all simple
undirected ER graphs with $N$ vertices and no self-edges
(i.e., networks with either zero or one edge between each
pair of distinct vertices). The partition function 
for the exponential networks reads as 
\be
Z=\sum_{G} e^{-H(G)}
\ee
where the measure in the sum over all graphs involves
the chemical potentials for some types of motifs, that 
is some kind of subgraph. 
For the 2-star model  the Hamiltonian reads as
\be
H =\theta L(G)-\mu s(G),
\label{01}
\ee
where $L(G)$ is the number of links and $s(G)$  is the number of triads.
For the Strauss model we have 
\be
H =\theta L(G)-\nu T(G),
\label{01}
\ee
where $T(G)$ is the number of 3 -cycles.

In ER network the links are independent with probability $p$, that is $L=\disp \frac{pN(N-1)}{2}$.
The mean-field solution of 2-star model   yields the consistency equation for the control
parameters
\be
p=\frac{1}{2}\disp \left[\tanh(\mu Np-\frac{1}{2}\theta) +1\right]
\label{crit1}
\ee
There is the critical line in the model $\theta=\mu N$ corresponding to the
first order phase transition which terminates at the second order 
transition critical point $\mu N=1$.

To proceed exponential random model with numeric, we use  the dynamic algorithm which replaces the numerical evaluation of the combinatorial problem \eq{01} by running stochastic evolution of network (the discrete Langevin dynamics), starting from some initial configuration until the evolution converges. The initial state of the network is prepared by connecting any randomly taken pair of vertices with the probability $p$  and the total number of links is fixed. Then, one randomly chooses  arbitrary link, say, between vertices $i$ and $j$ ($i,j$) and switch it to some other pair $(k, l)$ ($k \neq i,j;~l \neq i,j$). The acceptance probability this elementary permutation is defined by the standard Metropolis rule.
According to the graph Hamiltonian, if under the elementary step the number of triads is
increased or remains unchanged, a move is accepted, and if the number of triads is decreased by $\Delta$, a
move is accepted with the probability $e^{-\mu \Delta}$. The Metropolis algorithm runs repetitively for a large set of randomly chosen pairs of links, until it converges.

The Hamiltonian can be written in the form:
\be
H(G)=\theta \sum_{i<j}a_{i,j}-\mu \sum_{i<j<k}a_{ij}a_{jk},
\label{02}
\ee
where $a_{ij} =a_{ji}$ is an element of the adjacency matrix having value $1$ if an edge exists between vertices $i$ and $j$ and $0$ otherwise. If we consider the complement graph $\tilde{G}$ ($b_{ij} =1-a_{ji}$), the Hamiltonian \eq{02} keeps the same form:
\be
H(\tilde{G})\sim \alpha \sum_{i<j}b_{i,j}-\mu \sum_{i<j<k}b_{ij}b_{jk},
\label{03}
\ee
where $\alpha$ is linear combination of $\theta, \mu$ and the number of the vertices $N$.

\begin{figure}[ht]
\centerline{\includegraphics[width=16cm]{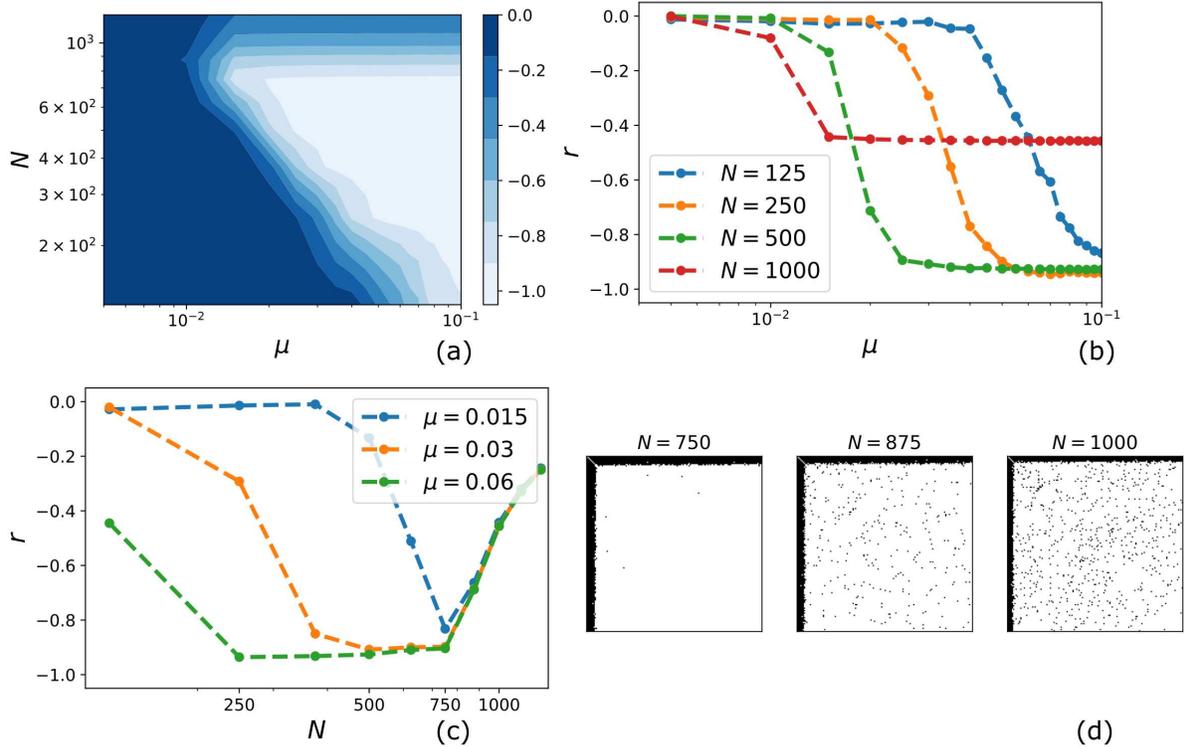}}
\caption{ Phase portrait of 2-star model  in $(\mu, N)$-plane for random graphs of $p=0.1$:
(a) density plot of the degree assortativity coefficient $r(\mu, N)$;
(b) the assortativity in dependence on the chemical potential $\mu$ for different network sizes;
(c) the assortativity in dependence on the network size $N$ for different value of $\mu$;
 (d) typical adjacency matrices for different network size and $\mu=0.1$.  The results are performed over averaging of 100 realization. }
\label{f01}
\end{figure}

\begin{figure}[ht]
\centerline{\includegraphics[width=16cm]{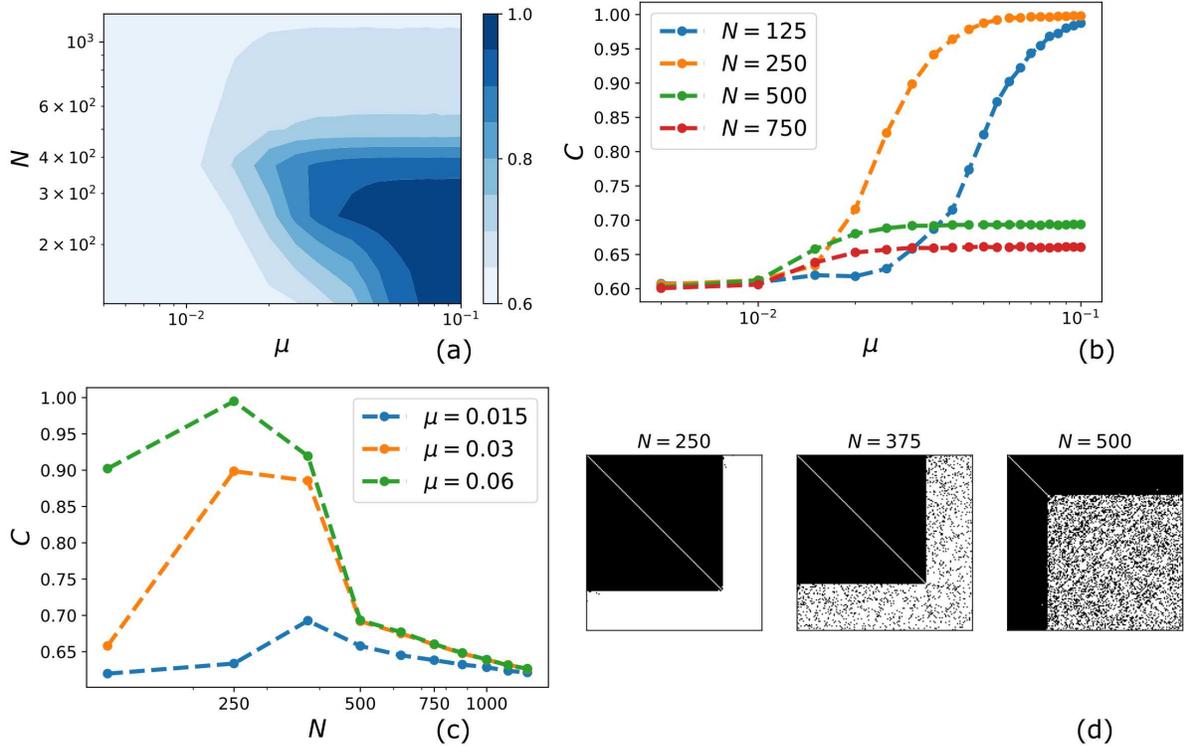}}
\caption{ Phase portrait of 2-star model  in $(\mu, N)$-plane for random graphs of $p=0.6$:
(a) density plot of the clustering coefficient $C(\mu, N)$;
(b) the clustering coefficient in dependence on the chemical potential $\mu$ for different network sizes;
(c) the clustering coefficient in dependence on the network size $N$ for different value of $\mu$;
 (d) typical adjacency matrices for different network size and $\mu=0.1$.  The results are performed over averaging of 100 realization. }
\label{f02}
\end{figure}

In 2-star model \cite{2star} there is the phase transition at some critical $\mu_{cr}$ 
and at $\mu>\mu_{cr}$ a kind of the dense component gets emerged which depends 
on the mean vertex degree $pN$. 
Hence similar  phase transition with the complete graph formation occurs for the complement graph $\tilde{G}$. 
The high-density network state described in \cite{2star} is a complete graph, for the fixed number of vertices the complement graph of this state has a star-like configuration (see \fig{f01}(d)). We compare analytically the ground state energies of two optimal configurations and found the following picture.  If the link probability $p<0.5$ the optimal configuration is a fat star-like graph (``fat star''), if $p>0.5$the ground state is a complete graph, $p_1=0.5$ is a natural critical point (see \fig{f01}(d), \fig{f02}(d)). 
In the Strauss model at $\nu>\nu_{cr}$ the single complete graph gets emerged \cite{strauss}.

The thermodynamic limit $N\rightarrow \infty$ for the 2-star and Strauss models  has been considered 
in the mean-field or graphon approaches \cite{old1,old2, alba}. It was claimed that in the thermodynamic
limit $\mu_{cr}\propto N^{-1}$ and the  models tend to the ER networks with rescaled probabilities $p^{*}$. 
 
\subsection{Crossover behavior at finite $N$}

We have revisited numerically the transition to the thermodynamical limit 
in 2-star and Strauss model. Contrary to the general expectations it turned out that the 
transition is not smooth and there is the crossover value $N^{*}$ where the
ground state of network gets changed. 
The typical phase portrait of the model in the parameter $(\mu, N)$ plane is shown in \fig{01}. 
For $p<0.5$ it is convenient to calculate the degree  assortativity coefficient which is defined as the Pearson correlation coefficient of degree between pairs of linked nodes:
\be
r=\frac{L\sum_{i=1}^{L}j_ik_i-\left[\sum_{i=1}^{L}j_i\right]^2}{L\sum_{i=1}^{L}j_i^2-\left[\sum_{i=1}^{L}j_i\right]^2},
\label{04}
\ee
where $L$ is the number of edges, $j_i, k_i$ are the degrees of the nodes, connected by the link $i$.
If the nodes of large degree are connected with nodes with a small number of links, the assortativity is close to $-1$, random graphs have the zero assortativity. The ground
state for $N<N^{*}$ gets identified with the dissortative  star-like state (``fat star'') while 
at $N>N^{*}$ we observe the smaller 
star-like graph with few hubs in the ER environment. The critical size is $N^{*} \approx 750$ for the network density $p=0.1$. Let us emphasize that it differs from the mean-field 
solution when the ER network without any defects was predicted. In the mean-field solution it is assumed
that the mean degree of the nodes is homogeneous and correspond to the dense cluster or ER phases.
Our numerical analysis demonstrates that there is the phase at $N>N^{*}$ when the ground state involves at least 
two components with different degrees: few linked hubs  and the ER vapor around.

For $p>0.5$ we observe the same crossover behavior, but from the clustered network state at $N<N^{*}$ to the smaller 
star-like graph with few hubs in the ER environment $N>N_{cr}$. The respective phase portrait is shown in \fig{02}. 

If we fix the value of $\mu$ a bit above critical point it is possible to estimate the $N^{*}(p)$
dependence, substituting in \eq{crit1} the critical line  equation $\theta=\mu N$:  
\be
N^{*}=\frac{\ln p- \ln (1-p)}{\mu(2p-1)}
\ee

In the Strauss model the crossover does exist as well. At $p=0.1$ and at $N<750$ the 
critical value of chemical potential behaves as $\nu_{cr}\propto 1/N$ and 
the ground state above the phase transition is identified with the full graph.
However at $p=0.1$ and at $N>750$ there is crossover regime when the complete graph 
starts to evaporate and
the ground state looks as the single hub in the ER environment. 

\begin{figure}[ht]
\centerline{\includegraphics[width=12cm]{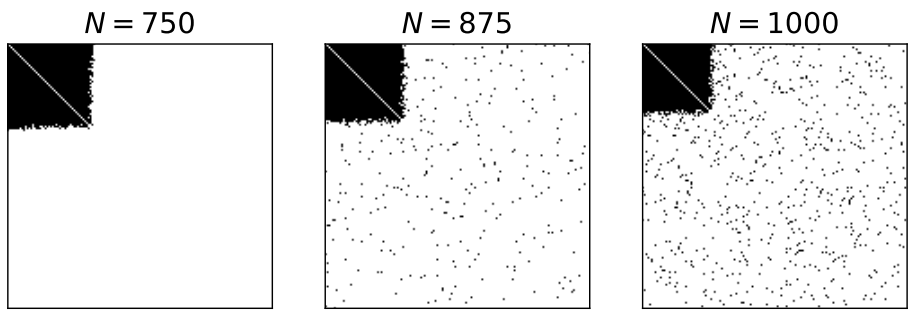}}
\caption{ Typical adjacency matrices of the Strauss model for different network sizes above the critical value of chemical potential ($\nu=1.0$). }
\label{f03}
\end{figure}

\subsection{Comments on the spectral behavior}

The spectral density is a good indicator of the 
the ground state of the network. There are some general
principles governing the evolution of the spectral
density as a function of chemical potential. First,
usually the spectral density in the continuum part 
of the spectrum gets modified. Secondly the isolated 
eigenvalues appear at the clusterization phase transition
since the number of the isolated eigenvalues corresponds
to the number of clusters. The creation of the cluster
has been identified as the eigenvalue instanton in \cite{crit}.

\begin{figure}[ht]
\centerline{\includegraphics[width=16cm]{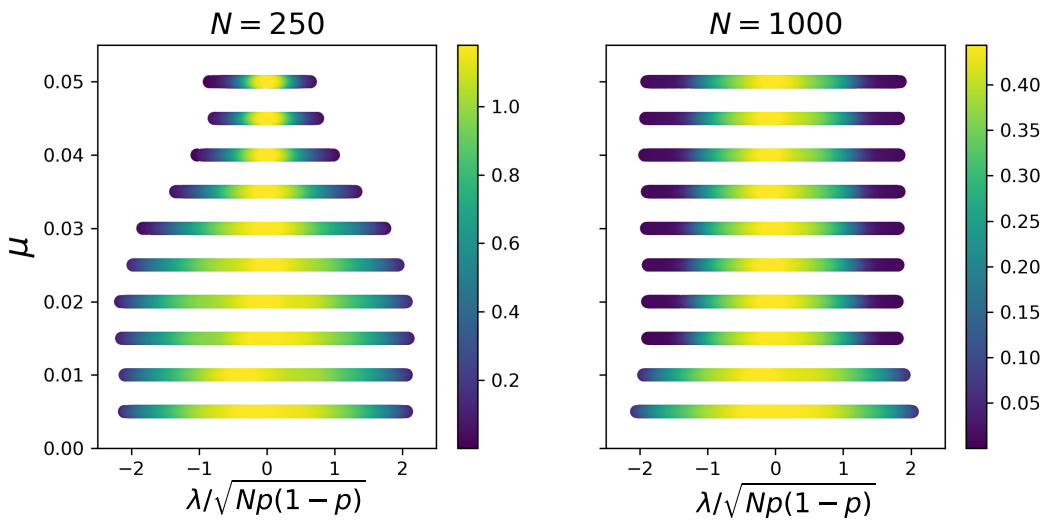}}
\caption{ Spectral behavior of the 2-star network below
and above the crossover. The results are performed over averaging of 100 realization ($p=0.1$).
}
\label{f04}
\end{figure}

Can we recognize the crossover behavior in the spectral
evolution? The answer turns out to be positive and is 
pictured at \fig{f04}. We investigated the dependence
of the spectral density on the chemical potential 
below and above $N^*$. At $N<N^{*}$ the spectral density 
becomes more  narrow at the phase transition point. 
On the other hand at $N>N^{*}$ the behavior is more rich.
First, as result of evaporation  the isolated eigenvalues get appeared symmetrically
with respect to the central zone. Since all hubs are 
identical in the ``fat star'', the eigenvalues at fixed
network are degenerate. However since the number of links
is different in each ensemble they form two zones which
are analogue of the non-perturbative zone found in \cite{crit}.
Increasing $N$ we increase the number of evaporated nodes which
results into the interaction between the star nodes and lifting the
degeneration of the eigenvalues.

\section{On the microscopic mechanism for crossover}
\subsection{Localization and entropic trap}

Let us discuss the possible microscopic mechanisms responsible 
for the crossover behavior. It is natural to assume that 
the mechanism of the cluster evaporation is of the entropic nature.
Indeed, due to  non-homogeneity of  network 
the entropic forces are present and could "`keep the nodes"
around some center of disorder if the entropic force is strong enough. 
Hence we could look for the examples
of the localization of trajectories around some center
of disorder at the random graph. We could mention two
relevant examples.

First, strong finite-size effects and crossover
behavior at some N have been found in the Anderson model at the regular
random graph \cite{mirlin1,biroli,mirlin2}. The crossover size $N_{cr}$ has been
interpreted as the existence of hidden intrinsic localization scale in the model.
It was observed the if $N<N_{cr}$ the system behaves almost as critical even near the critical
line while at $N>N_{cr}$ the RG flow pushes the system in the different direction
to the ergodic state. This is similar to what we have found: at small N we have the
critical ``fat star'' or cluster even apart from the critical line while at 
$N>N_{cr}$ they start to evaporate down to ergodic ER network with small
inhomogeneity.

Another interesting possibility for the crossover mechanism can
be traced from the heavy root problem \cite{tamm}. In that 
paper the infinite tree-like network with equal node degrees d and single node with large degree $d_0$
has been  considered. It was demonstrated that  depending
on a degree of the heavy root the random walks are localized 
around the heavy root or escape to infinity. The result has been
derived analytically and confirmed numerically. The critical
relation reads as 
\be
d_{0,cr}=d^2-d
\ee
The localization of the random walks occurs at $d_0>d_{0,cr}$ and
at criticality the interesting thermal-like behavior was reported.

Let us compare this result with the crossover in the 2-star model.
We have found that at $N>>N^{*}$ the ground state is the single hub and let us interpret it
as the version of the heavy root problem. In such interpretation
$d_0=N$ while we can estimate $d=pN$ for the ER environment. Hence
the condition of the entropic trapping of the trajectories around the hub is
\be
N>(pN)^2
\ee
which can be fulfilled only at $N< 1/p^2$. Of course this estimate
is rough enough but it demonstrates that ``fat star'' ground state can exist
only below some critical value of N. This is realization of the
entropic trap mechanism for crossover in 2-star model. To clarify
this scenario better it is necessary to analyze the heavy root
problem for the ER network.

\subsection{Gravity analogy}

It is well-known that large N matrix model was used for the effective
summation over 2d geometries (see \cite{ginsparg} for review). The partition function of the matrix model
is presented in the following form
\be
Z(t_k)=\int DM exp(\sum_k t_kTrM^k)
\ee 
where $t_k$ are the chemical potentials for single trace observables and M belongs to some 
matrix ensembles. The rich critical behavior of the partition 
function corresponds to the criticality of the pure topological 2d gravity if only $t_2,t_3$ 
are non-vanishing or topological gravity coupled to matter in the general 
case of multiple chemical potentials. The analogy with the exponential graph
partition function is clear. In that case $t_k$ counts the number of k-cycles
in the network and for instance the critical behavior in the Strauss model
corresponds to the critical behavior in the pure topological 2d gravity.
The Laplacian matrix of the random network  $L= D-A $, where 
D is diagonal matrix of the mode degrees $D = diag (d_1,\dots d_N)$ and A is a adjacency matrix, 
is the discretization of the 2d Laplace
operator on the Riemann surface.

The size of the matrix plays the role
of the inverse coupling constant in the quantum gravity $g_s=1/N$
hence large N limit corresponds to the weak coupling regime and 
the finite-size effects correspond to the "`perturbative"' 
corrections. There
are also the non-perturbative effects with  $e^{-N}$ dependence
which are identified in matrix models as eigenvalue instantons 
or as the creation of the baby universes in the gravity language \cite{baby}.
These non-perturbative effects have been identified in the
constrained ER networks and RRG in \cite{crit} as the cluster formation. Our findings
in this study suggest that account of the finite-size effects $\frac{1}{N}$
strongly influence the critical behavior and
could lead to the crossover-like  regime in the matrix models. Note 
that in the 2-star model we take into account  both singlet and  the non-singlet 
sectors in matrix model language since the number of 2-stars
can not be presented as the single trace of the power of adjacency matrix of the graph.

Another viewpoint concerning the gravity interpretation of our 
results is suggested by the holographic approach for large N
boundary theories. 
The holographic setup provides the way to describe the dual BH
in terms of large number of D0 branes in the boundary theory \cite{cotler, hanada}.
We can consider our exponential random network as the toy  boundary 
large N theory.
The single  BH in the holographic dual looks as the full block in the large N matrix 
where the diagonal matrix elements represent the positions of the D0 branes
and the off-diagonal elements correspond to the strings connecting D0 branes.
The several blocks at the diagonal represent the configuration of the several 
black holes while
the emission of the single D0 brane from the black hole  looks as one separated element in the
matrix on the diagonal outside the cluster \cite{cotler, hanada} .

Could we learn some lessons from our findings for the toy matrix BH?
First note the the very creation of the cluster or equivalently BH
is the dynamical critical phenomena. It is induced by the proper value of the
chemical potential which makes the joining of the short strings preferable process.
This suggests the microscopic mechanism of the toy BH formation via 
phase transition. The next point concerns the evaporation of the toy BH which
we could relate with the evaporation of the cluster. Since the coupling constant behaves as 
$\frac{1}{N}$ increasing the network size we decrease the perturbative gravity effects 
and finally at some critical $N^{*}$ the BH starts to evaporate since 
"`particles" can escape. Certainly
this picture is quite qualitative but it could shed some light at the entropic
mechanism behind the BH formation and evaporation.

The BH analogy is consistent with the embedding of the random network
into the AdS hyperbolic geometry \cite{krioukov}. The key point is the clear relation between
the value
of the radial coordinate in AdS and the degree of nodes \cite{krioukov}. The region near the center
of AdS corresponds to the nodes with the highest degrees. The block of a matrix X
representing the toy matrix BH  corresponds to the full graph hence it
involves
the nodes with the same degrees if we identify X with the
Laplacian matrix or the graph. Therefore according the degree-radial coordinate duality 
all of them are at the same value of radial
coordinate which presumably can be identified with the position of BH horizon. This 
is also consistent  with the interpretation of the diagonal values of the matrix
in \cite{hanada,cotler} as the positions of D0 branes. When the evaporation 
of the full graph happens the nodes with smaller degrees get decoupled and
due to the degree-radial coordinate correspondence they occur at the smaller
value of the radial coordinate justifying the BH evaporation picture. We 
shall discuss the holographic interpretation of the embedding of the 
random network into hyperbolic geometry in more details elsewhere.

The evaporation 
of the ``fat star'' configuration in 2-star model has the 
similar interpretation with additional subtlety. Due to the "`open-string"' non-singlet
sector in the 2-star model there is a kind of hair in the ground state 
which presumably corresponds to the black hole hair. Indeed apart from the full graph
block which represents the horizon there are finite number of nodes with the smaller 
degrees which are placed at the different radial coordinate. Since they are connected
with the full graph core we have a kind of BH hair.
Upon the evaporation the ``fat star'' gets transforms into the single hub but 
the hair remains intact.

\section{Conclusion}

In this Letter we have found the strong impact of finite-size $1/N$ effects on  critical
behavior  of exponential random graphs. Both in 2-star and Strauss models there are the 
critical sizes $N^*(p)$ when the crossover takes place. The key
process at the crossover region is the evaporation of the ``fat star''
or the single cluster which represent the ground state at $N<N^{*}$. 
The final stage of the cluster evaporation in 2-star model
is the single complete star.

We have briefly discussed the possible microscopic explanation
of the crossover found but the further analysis is certainly required. 
It would be interesting to understand what kind of modification of the 
mean-field analysis if any is required to describe the
crossover regime. It is also very desirable to elaborate 
the analogy with the BH evaporation and localization
at heavy root in the random network environment further.

Two further directions are quite evident. First, it would 
be  interesting to extend our analysis of finite-size effects for the networks
with conserved degrees, for instance RRG or constrained
ER networks. In that case above the critical point we have 
multiple clusters and generically we have the cascade of the
phase transitions instead of the single one. Presumably we 
have the cascade of the evaporation of clusters. Since 
formation of clusters is nothing but the eigenvalue instanton phenomena
suppressed as $e^{-N}$ we have the interesting possibility
to elaborate numerically the analogue of non-perturbative
phenomena in the quantum gravity and interplay between the
perturbative and non-perturbative effects.

Another question concerns
the elaboration of the finite-size effects in the multicolor 
exponential graphs. The corresponding analysis will be
presented elsewhere \cite{gv}. It would be also interesting
to  investigate the spectral statistics as function of N which
would answer the question concerning the localization
properties of the modes.

We would like to thank A. Andreev and S. Nechaev for the useful discussions.
The work of A.G. is partially supported by Basis Foundation Fellowship and 
grant RFBR 19-02-00214. The work of O.V. is partially supported by Basis Foundation Fellowship and 
grant RFBR 18-29-0316.

\end{document}